\documentclass{ws-procs975x65}            % default, citations in superscript
\begin{document}

\begin{flushright}
{\small LLNL-PROC-676444}
\end{flushright}

\title{Lattice study of the scalar and baryon spectra in many-flavor QCD}

\author{
Yasumichi Aoki$^a$, Tatsumi Aoyama$^a$, Ed Bennett$^b$, 
Masafumi Kurachi$^c$, 
Toshihide Maskawa$^a$,  Kohtaroh Miura$^d$, 
Kei-ichi Nagai$^a$, Hiroshi Ohki$^e$, 
Enrico Rinaldi$^f$, Akihiro Shibata$^g$, 
Koichi Yamawaki$^a$, and Takeshi Yamazaki$^h$ \\
\centering{(LatKMI Collaboration)}
}

\address{
$^a$ 
Kobayashi-Maskawa Institute for the Origin of Particles and the Universe (KMI), 
Nagoya University, Nagoya, 464-8602, Japan \\
$^b$ 
Department of Physics, Swansea University, Singleton Park, 
Swansea SA2 8PP, UK \\
$^c$ 
Institute of Particle and Nuclear studies, High Energy Accelerator Research
Organization (KEK), 
Tsukuba 305-0801, Japan \\
$^d$ 
Centre de Physique Theorique(CPT),
Aix-Marseille Univerisity, Campus de Luminy, Case 907,
163 Avenue de Luminy, 13288 Marseille cedex 9, France \\
$^e$ 
RIKEN/BNL Research center, Brookhaven National Laboratory, 
Upton, NY, 11973, USA \\ 
$^f$ 
Lawrence Livermore National Laboratory, Livermore, 
California, 94550, USA \\
$^g$ 
Computing Research Center, High Energy Accelerator Research Organization (KEK),
Tsukuba 305-0801, Japan \\
$^h$ 
Graduate School of Pure and Applied Sciences,
University of Tsukuba, Tsukuba, Ibaraki 305-8571, Japan \\
}

\begin{abstract}
In the search for a composite Higgs boson in walking technicolor models, many flavor QCD, 
in particular with $N_f=8$, is an attractive candidate, 
and has been found to have a composite flavor-singlet scalar as light as the pion. 
Based on lattice simulations of this theory with the HISQ action, we will present our preliminary results 
on the scalar decay constant using the fermionic bilinear operator, 
and on the mass of the lightest baryon state 
which could be a dark matter candidate. 
Combining these two results, 
implications for dark matter direct detection 
are also discussed.
\end{abstract}

\keywords{
Lattice gauge theory, Conformal/Walking Dynamics, 
Dark matter, composite Higgs particle.}

\bodymatter

\section{Introduction}\label{intro}
A new scalar particle with mass $125$ GeV has been discovered at the LHC, 
and the current experimental data is consistent with 
the standard model (SM) of particle physics. 
However, the origin of mass and electroweak symmetry breaking remains unknown, 
and the SM does not have a viable dark matter (DM) candidate,
which are reasons why we are seeking new physics beyond the SM. 

It is possible that there is strong gauge dynamics beyond the TeV scale: 
one popular candidate is the 
Walking
Technicolor model. This breaks 
the electroweak symmetry 
dynamically  
in an approximately scale-invariant 
dynamics, and predicts
a large anomalous dimension $\gamma_m\simeq 1$ 
together with a ``techni-dilaton'' as a 
pseudo Nambu-Goldstone boson 
of the spontaneously broken approximate scale symmetry\cite{Yamawaki:1985zg}.
The Higgs scalar could 
be identified with the techni-dilaton, 
which is a composite bound-state of the flavor-singlet techni-fermion bilinear,  
parametrically lighter than 
other techni-hadrons. 
From a phenomenological point of view, 
the techni-dilaton decay constant (denoted here as $F_\sigma$) 
is a 
very important parameter in addition to its mass, 
since $F_\sigma$ controls all the techni-dilaton's couplings to SM particles. 
In addition, the technicolor model has 
a 
rich hadron structure,
and composite resonances like the techni-rho meson ($\rho$) 
may be in the discovery reach of run-II at the LHC.
Furthermore the techni-baryon, its lightest neutral component, 
could be a stable particle 
due to techni-fermion number conservation.
Thus it would be a candidate for DM. 

In order to study the walking technicolor model, 
non-perturbative understanding of the strong gauge dynamics
is necessary, and  
lattice numerical calculations are the most powerful tool 
for that purpose.
In a previous lattice study, the LatKMI collaboration has shown 
that 8-flavor ($N_f=8$) QCD could be a candidate 
for a walking gauge theory. 
We found an approximate hyperscaling relation for various hadron masses 
in a certain fermion mass range\cite{Aoki:2013xza}. 
Remarkably, we also found that the flavor singlet-scalar ($\sigma$) mass 
is as light as the pseudoscalar ($\pi$), 
which might indicate that $\sigma$ could be regarded as the techni-dilaton~\cite{Aoki:2014oha}.

In this proceedings 
we investigate the decay constant of the flavor-singlet scalar ant its mass.
Using the Ward-Takahashi (WT) relation of the scale symmetry 
in the continuum theory, 
we estimate the dilaton decay constant $F_\sigma$. 
As for the techni-baryon DM, it is important to investigate its direct detection,
where the most dominant contribution of the DM scattering amplitude
is the scalar mediated interaction.
Based on a DM 
effective
theory whose parameters include 
$F_\sigma$ and the techni-baryon mass, 
we evaluate the leading order of the scattering cross section 
of the DM. 
Using the lattice input for the above parameters, 
we discuss 
experimental detectability and the possible realization of the techni-baryon 
DM scenario. 

In the next section, we briefly explain how to calculate 
the mass of the flavor-singlet scalar and scalar decay constant on the lattice.
We discuss a relation between the scalar and dilaton decay constants.
In Section 3, we show our lattice result of the scalar mass and decay constants, 
including a chiral extrapolation based on a dilaton effective theory.
In Section 4, we study the DM detection rate based on the dilaton 
chiral perturbation theory with baryon, and the summary is given in Section 5.
Note that all the results shown in this proceeding are preliminary.

\vspace{-3mm}
\section{Scalar mass and decay constant}

We investigate the mass of the scalar and its decay constant 
on the lattice. 
The mass of the flavor-singlet scalar, which we call $\sigma$, 
is calculated from the two-point correlation function ($C_\sigma(t)$) 
of the flavor-singlet scalar bilinear operator $\mathcal{O}_s$.
By using the staggered lattice fermion field $\chi_i$, 
$\mathcal{O}_s$ is written as $\mathcal{O}_s(x, t) = \sum_i \bar{\chi}_i(x,t)\chi_i(x,t)$, 
where $i$ denotes the different staggered fermion species, namely $i=1, 2$ for $N_f=8$ QCD. 
The correlator is then given by
\begin{eqnarray}
C_\sigma(t) &=& \frac{1}{V} 
\sum_x 
\left\langle  \mathcal{O}_s(x,t)  
\mathcal{O}_s(0,0) \right\rangle, 
\label{eq:C}
\end{eqnarray}
where $V$ is the spatial volume ($V=L^3$).
The asymptotic behavior of $C_\sigma(t)$ 
is given by 
$C_\sigma(t)=
A_\sigma(t)+(-1)^t A_{\pi_{\overline{SC}}} (t)$,   
where $A_{\pi_{\overline{SC}}}$ is a pseudoscalar correlator,  
and $A_H(t)=A_H (e^{-m_H t}+ e^{-m_H(T-t)})$ for different hadrons $H$.

We define the scalar decay constant $F_S$ as  
the scalar operator matrix element, 
\vspace{-1mm}
\begin{eqnarray}
\langle 0 | m_f O_s(0,0) |\sigma(0) \rangle
=F_S m_\sigma^2.
\end{eqnarray}
\vspace{-1mm}
The above matrix element can be calculated from $C_\sigma(t)$.
Substituting the complete set, $\sum_n | n \rangle \langle n| 
= \int \frac{d^3p}{(2\pi)^3} \frac{|\sigma(p) \rangle \langle \sigma(p)|}{2E_p} +\cdots$
into Eq.~\ref{eq:C},  the correlator $C_\sigma(t)$ is given by
\vspace{-2mm}
\begin{eqnarray}
C_\sigma(t)= \frac{1}{V}|\langle 0|O_s(0,0) |\sigma(0) \rangle|^2 \frac{1}{2m_\sigma}
\left(e^{-m_\sigma t}+e^{-m_\sigma (T-t)}\right).
\end{eqnarray}
\vspace{-1mm}
Thus we calculate $F_S$  as follows,
\vspace{-1mm}
\begin{eqnarray}
F_S = \frac{m_f \sqrt{2m_\sigma VA_\sigma}}{m_\sigma^2}.
\label{eq:Fs}
\end{eqnarray}
We note that this quantity is renormalization group invariant and a physical quantity, 
and it is easy to measure
on the lattice.
It is also noted that $F_S$ should obey a hyperscaling relation, $F_S \propto m_f^{1/(1+\gamma)}$
with $\gamma\sim \gamma_m$,
if the theory is in the conformal window~\cite{DelDebbio:2010ze}.

Let us now discuss the dilaton decay constant $F_\sigma$.
The following discussion is based on the continuum theory.
It is important to recall the basic definition of the dilaton decay constant, 
which is $\langle 0 | \mathcal{D}^\mu(x)  |\sigma(p) \rangle =- i F_\sigma p^\mu e^{-ipx}$.
From this we obtain 
 $\langle 0 | \partial_\mu \mathcal{D}^\mu(0)  |\sigma(0) \rangle =- F_\sigma m_\sigma^2$. 
Therefore the dilaton decay constant can be directly calculated from the matrix element 
of the dilatation current. However the dilatation current is rather difficult to construct 
on the lattice, since it contains a power divergence that needs to be subtracted.

Instead, we consider an alternative way to estimate it,
 which 
can be derived by using the WT relation for the dilatation current. 
Following the argument in \cite{Bando:1986bg}, 
we consider the integrated WT relation for dilatation transformation. 
We use the scale transformation relation of an operator $\mathcal{O}$, 
$ \delta_D \mathcal{O}= \Delta_\mathcal{O} \mathcal{O}$, 
where $\Delta_\mathcal{O}$ is the scaling dimension of $\mathcal{O}$.
In the zero momentum transfer limit the WT relation leads to 
$\int d^4 x T \langle \partial_\mu \mathcal{D}^\mu(x) \mathcal{O}(0) \rangle
= \Delta_\mathcal{O} \langle \mathcal{O} \rangle$\footnote{
In the formula, 
the vacuum contribution on the left hand side is absent in the WT relation. 
}, from which, by assuming  the $\sigma$ pole dominance,  we obtain
$ \Delta_\mathcal{O} \langle \mathcal{O} \rangle= (-F_\sigma m_\sigma^2)\frac{1}{m_\sigma^2} \langle \sigma |\mathcal{O}|0 \rangle=-F_\sigma \langle \sigma |\mathcal{O}|0 \rangle$.\footnote{
The so-called Partially Conserved Dilatation Current (PCDC) relation 
$F_\sigma^2 m_\sigma^2 =-  4  \langle \theta_\mu^\mu\rangle$ follows for $\mathcal{O}= \partial_\mu \mathcal{D}^\mu=\theta_\mu^\mu$ (nonperturbative trace anomaly).} 
In the
case we take $\mathcal{O}(x)=m_f \sum_i^{N_F} 
\bar\psi_i \psi_i (x)$ 
(the flavor-singlet scalar operator), 
we obtain a relation, 
\begin{eqnarray}
F_S F_\sigma m_\sigma^2=- \Delta_{\bar{\psi}\psi} m_f 
\sum_i^{N_F}
\left\langle  \bar\psi_i \psi_i \right\rangle.
\label{eq:Fs}
\end{eqnarray}
 (Dividing both sides 
 by $m_f$ leads to the relation obtained at $m_f=0$ \cite{Matsuzaki:2012xx}.)

We note that this relation 
holds in the continuum theory 
with infrared conformality under the assumption 
of the $\sigma$ pole dominance. 
While we do not know 
the
real infrared behavior in $N_f=8$ QCD towards 
the 
chiral limit, 
our recent lattice study shows that an approximate 
hyperscaling relation (walking behavior) 
is found for various hadron masses
in a certain fermion mass region of $m_f$, 
so the above relation could be effective also in the walking region 
up to excited state contaminations and discretization effects.
Therefore in the following preliminary lattice analysis, 
we shall use this relation for a semi-direct estimate of $F_\sigma$. 
In order to precisely calculate $F_\sigma$, however, 
a direct calculation of the dilatation current is needed.

\vspace{-4.5mm}
\section{Result}

\begin{figure}[h]
\begin{center}
\includegraphics[width=1.8in,clip]{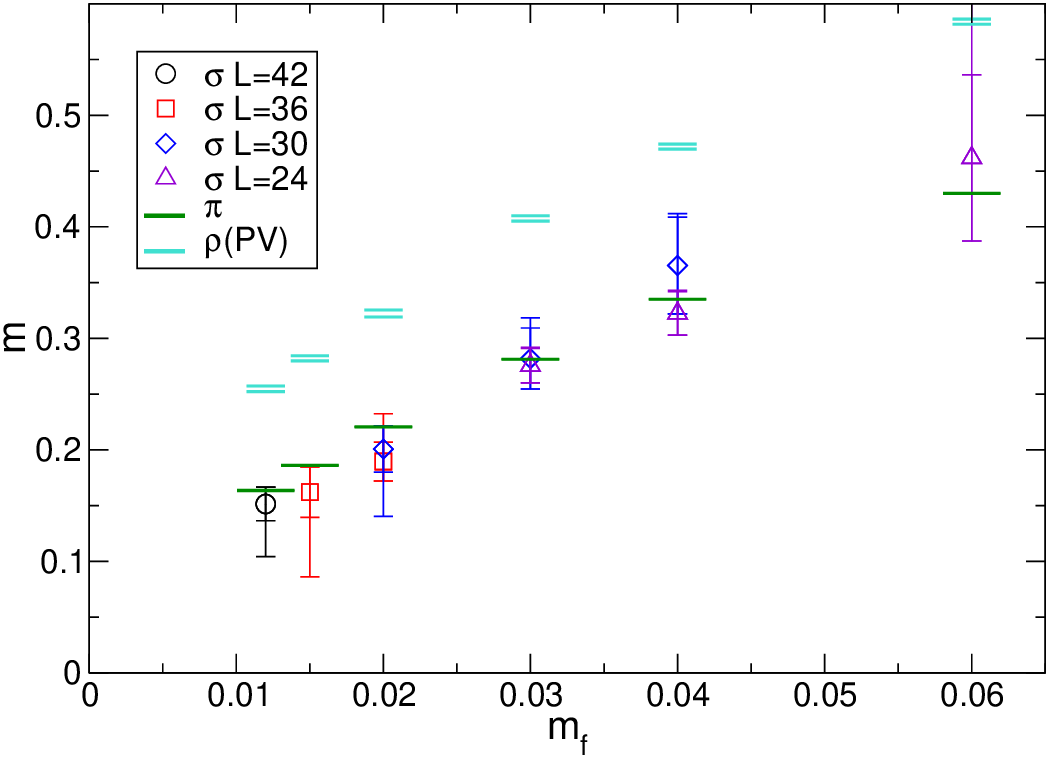}
\hspace{5mm}
\includegraphics[width=1.8in]{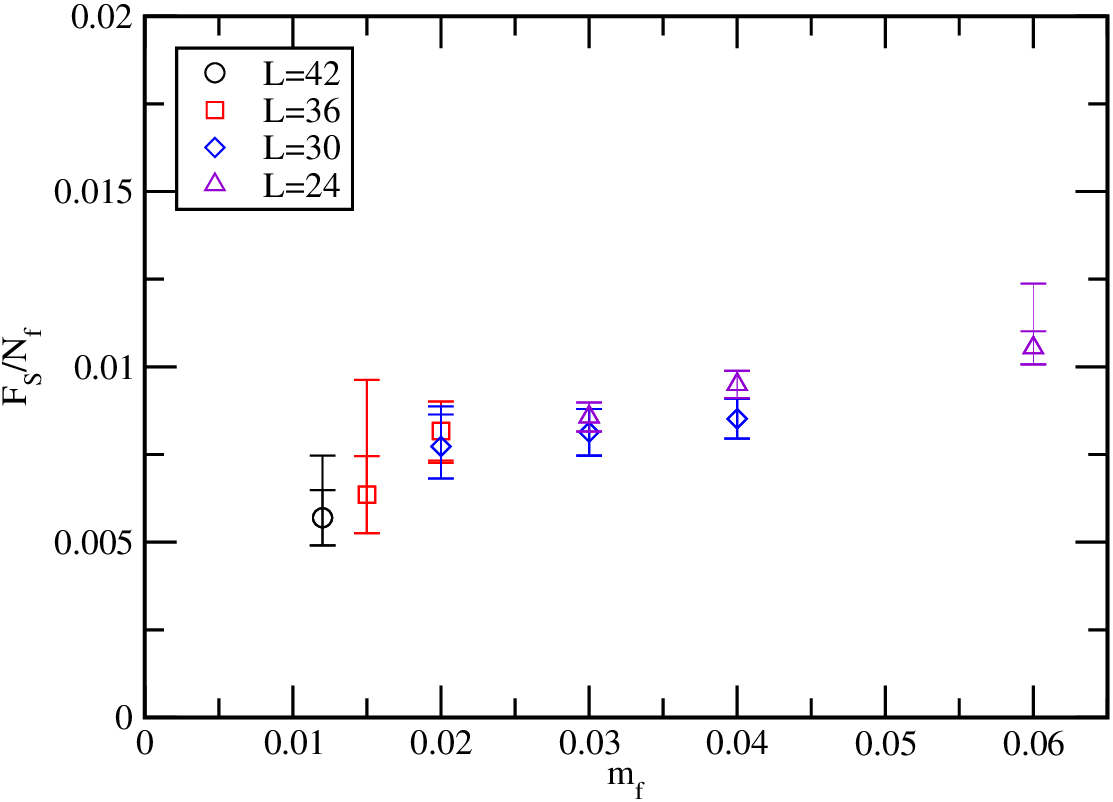}
\end{center}
\caption{
(Left) Mass of the flavor-singlet scalar.
Other hadron masses are also shown. 
Outer error represents the statistical and systematic uncertainties added in quadrature, 
while inner error is only statistical.
(Right) Scalar decay constant of the flavor-singlet scalar.
}
\label{fig:m_sigma}
\end{figure}

In this section we show our lattice results for $N_f=8$ QCD. 
Details of the simulation and analysis for measurements 
can be found in \cite{latkmi}.
The results for $m_\sigma$ and $F_S$ are summarized in Fig~\ref{fig:m_sigma}.
It is found that the scalar is as light as $\pi$,
and clearly lighter than $\rho$, 
and we obtain a good signal for $F_S$ as well. 
The lightness of the scalar could be regarded as 
a reflection of a dilatonic nature of the scalar. 

\begin{figure}[h]
\begin{center}
\includegraphics[width=1.8in]{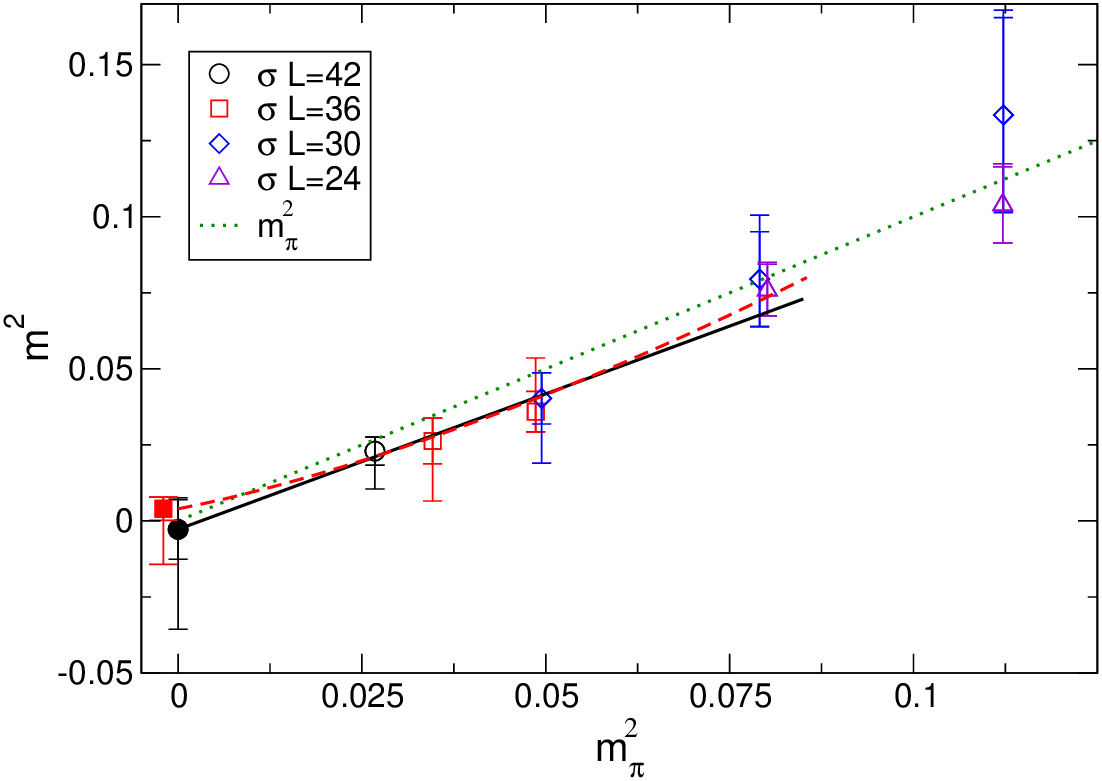}
\hspace{5mm}
\includegraphics[width=1.8in]{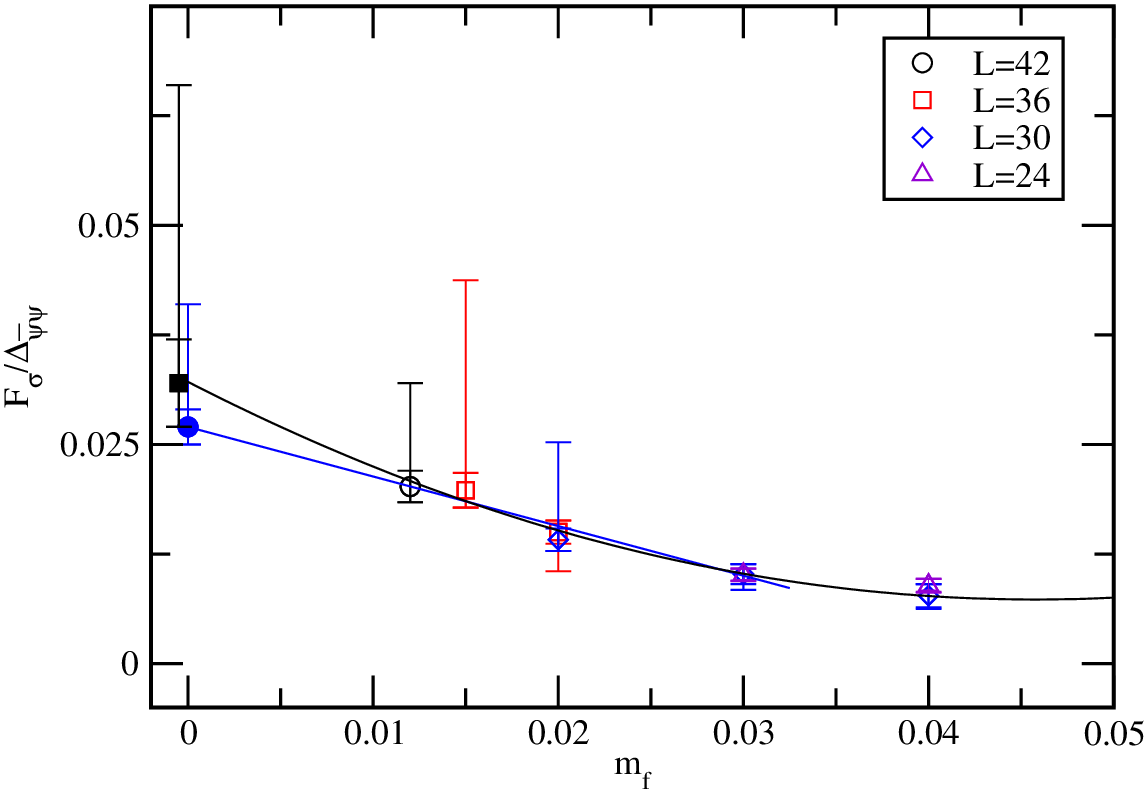}
\end{center}
\caption{
(Left) $m_\sigma^2$ v.s. $m_\pi^2$. 
Black line shows the fit with DChPT formula.
Dashed red curve shows the fit result with a naive form, $m_\sigma=c_0 +c_1 m_f$.  
(Right) Dilaton decay constant from a semi-direct method.
Blue and Black lines show the chiral fits
of the linear and quadratic polynomial in $m_f$ 
with the lightest 4 and 5 data points used.
Outer error represents the statistical and systematic uncertainties added in quadrature, 
while inner error is only statistical.
}
\label{fig:chiral}
\end{figure}

While our fermion mass is far from the chiral limit, 
we shall study the chiral extrapolation based on an effective theory, 
which is dilaton chiral perturbation theory (DChPT) \cite{Matsuzaki:2013eva}.  
At the leading order of the DChPT, 
the scalar mass is given by $m_\sigma^2 = d_0 +d_1 m_\pi^2$, 
where $d_1 = \frac{(3-\gamma_m)(1+\gamma_m)}{4}\frac{N_fF_\pi^2}{F_\sigma^2}$, 
$m_\pi$ and $F_\pi$ are the mass and decay constant of $\pi$. 
We also try a fit with a naive form of $m_\sigma = c_0+c_1 m_f$.
The fit results are shown in the left panel of Fig.~\ref{fig:chiral}, 
where a reasonable value of $\chi^2/$dof $\sim \mathcal{O}(1)$ is obtained.
The value of $m_\sigma^2$ in the chiral limit is $d_0=-0.0028(98)$. 
Thus we obtain a very light $\sigma$. 
From the fit value of the slope $d_1=0.89(26)$ in DChPT, 
we also obtain $F_\sigma \sim \sqrt{N_f}F_\pi \sim 0.06$,
with $F_\pi$ being in the chiral limit \cite{latkmi} and $\gamma\sim \gamma_m  \sim 1$. 

On the right panel of Fig.~\ref{fig:chiral}, 
the result of $F_\sigma$ from the semi-direct estimate (Eq.~\ref{eq:Fs}) 
is shown\footnote{For the chiral condensate, we use its chiral limit value to avoid 
large lattice artefacts.}. 
We also carry out the chiral extrapolation fits, whose 
results are also shown in the figure.
In the chiral limit, we obtain $\frac{F_\sigma}{\Delta_{\bar\psi \psi}} \sim 0.03$.
Given that $\Delta_{\bar\psi \psi} =3-\gamma$,  
we have shown how two different methods, DChPT and a semi-direct calculation, 
give a consistent result for $F_\sigma$. 

\vspace{-2mm}

\vspace{-2mm}

\section{Dilaton ChPT with baryon and dark matter}

The technicolor model may provide a good candidate 
for composite DM, in the form of 
a neutral baryonic bound state made of 
constituent (possibly charged) techni-fermions.
In this case, the coupling between the SM particle and the DM 
as well as the mass of the DM are constrained 
by direct detection experiments. 
In most cases, 
the experiments determine the scattering rate 
of DM with heavy nuclei in the detector, 
one dominant contribution to which is 
the Higgs (scalar) mediated spin-independent process.
In the following, we provide 
a low-energy effective theory of the walking gauge theory 
including the DM, whose low-energy constants can be determined from 
the lattice results of the baryon and scalar spectra.
The result obtained in the previous section is useful, 
and it gives rise to 
information for the DM direct detection experiments.

We consider a DM effective theory including 
the dilaton based on the DChPT. 
Since the DM is the lightest techni-baryon, 
the extension to the baryon sector of the DChPT is 
straightforward \cite{Matsuzaki:2013eva}.
As a result, in the leading order  
the dilaton field can only couple through the baryon mass term as 
\begin{eqnarray}
\mathcal{L} = \bar{B}(x)(i\gamma_\mu \partial^\mu - \chi(x) m_B )B(x),
\end{eqnarray}
where $\chi(x) = e^{\sigma(x)/F_\sigma}$, $B(x)$ is 
the baryon field,  and $m_B$ is its mass in the chiral limit.
The parameter $m_B$ explicity breaks the scale symmetry, 
and the (pseudo) dilaton acts on this term to 
make the action 
scale invariant.
Actually this action is invariant under the scale transformation, 
$\delta B = (\frac{3}{2}+x_\nu \partial^\nu)B$, $\delta \chi = (1+x_\nu \partial^\nu)\chi$.

From this effective theory, we can read off 
the dilaton-baryon effective coupling ($y_{\bar{B}B\sigma}$),  
which is uniquely determined as 
$y_{\bar{B}B\sigma}=m_B/F_\sigma$.
Regarding to the SM sector, 
the dilaton-nucleon coupling is also determined 
within the framework of the dilaton effective theory~\cite{Matsuzaki:2012vc}, 
since the dilaton-quark coupling ($y_{\sigma \bar{f} f}$) could be 
related to the SM Yukawa coupling ($y_{h_{\mathrm{SM}}\bar{f}f}$) as 
$\frac{y_{\sigma\bar{f}f}}{y_{h_{\mathrm{SM}}\bar{f}f}}=\frac{(3-\gamma)v_{EW}}{F_\sigma}$.
Combining both SM and technicolor sectors, 
the cross section with a target nucleus $N$ 
is given as 
$\sigma_{SI} = \frac{M_R^2}{\pi}(Z f_p+(A-Z)f_n)^2$,
where $M_R=(m_B m_N)/(m_B+m_N)$, 
and 
$Z$ and $A$ are the total number of the 
protons ($p$) and neutrons ($n$) in the nucleus. 
The parameter $f_{(n,p)}$ is defined as 
$f_{(n, p)}=\frac{m_N}{\sqrt{2} m_\sigma^2} \frac{y_{\bar{B}B\sigma}}{F_\sigma}
(3-\gamma)\left( \sum_{q=u,d,s} f_{T_q}^{(n,p)}+\frac{2}{9}f_{T_G}^{(n,p)}  \right)$,
where $f_{T_q}^{(n,p)}$ is the nucleon $\sigma$-term 
of the light quarks ($q=u,d,s$), and $f_{T_G}^{(n,p)}$ is that of the 
heavy quarks\footnote{
In general, techni-fermions can be charged under the SM color, 
so there may exist additional contributions to the nucleon matrix elements
from the techni-fermions. 
In this analysis, we omit these contributions for simplicity. }.
Thus lattice calculations are used in the technicolor theory as well as 
in QCD theory to obtain non-perturbative information about DM physics.

\begin{figure}[h]
\begin{center}
\includegraphics[width=2in,clip]{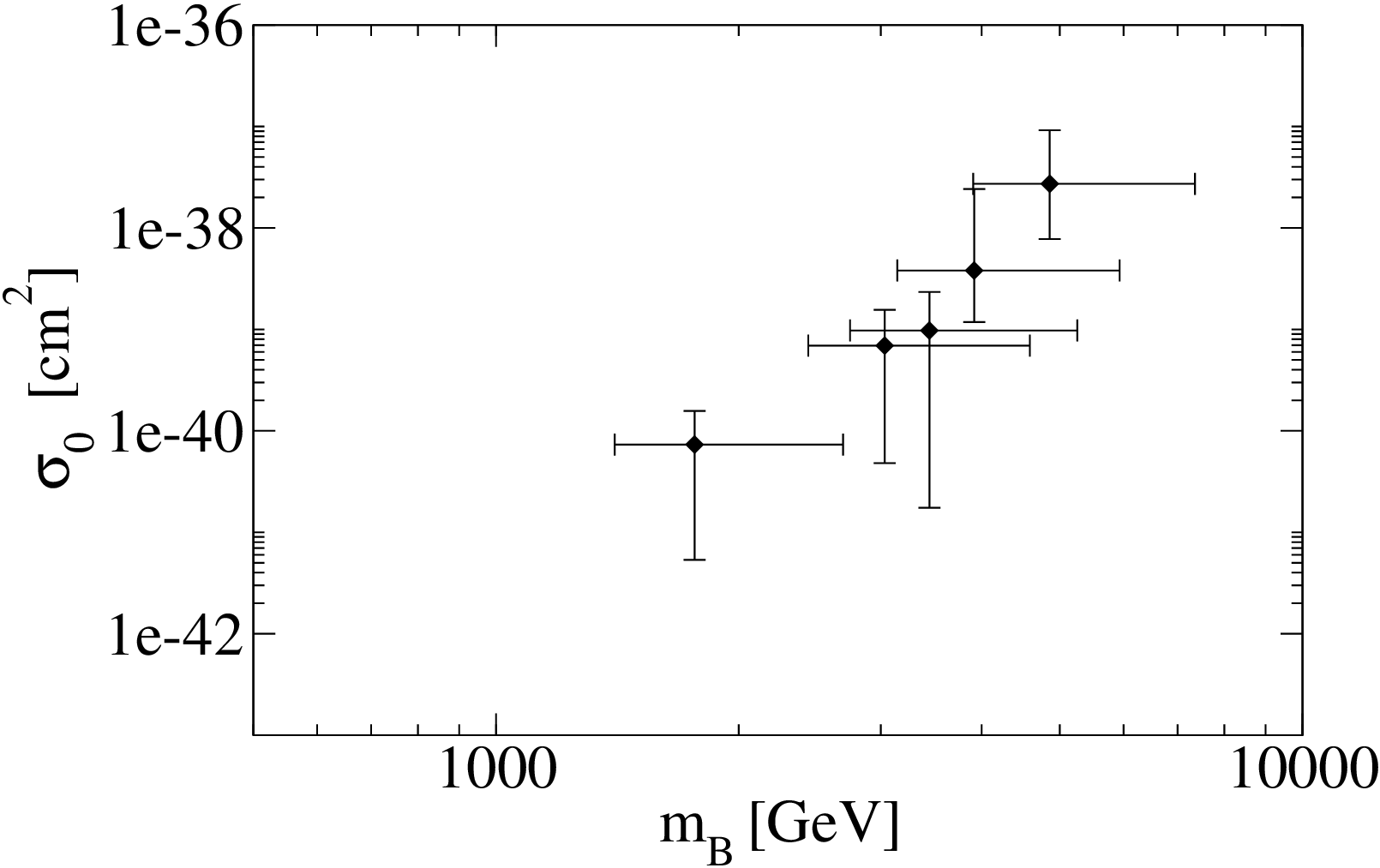}
\end{center}
\vspace{-3mm}
\caption{
$\sigma_0$ [cm$^2$] as a function of $m_B$ [GeV]. 
The results for $m_f=0.030, 0.020, 0.015, 0.012$, and 
the chiral limit from upper-right to lower-left.
Both the statistical and systematic errors are included.
Experimentally allowed region is below the plotted window.
}
\label{fig:sigma0}
\end{figure}

Here we show our numerical results of 
the DM cross section\footnote{ 
A similar analysis on the lattice has been performed for a different composite DM model 
based on strong dynamics~\cite{Appelquist:2014jch}.}.
We use the lattice results of the dilaton decay constant $(F_\sigma)$ 
obtained from the previous section and baryon mass, 
while the scalar mass $m_\sigma$ is fixed to its experimental value 
($125$ GeV) in this analysis. 
We use the values in \cite{Hisano:2015rsa} for $f_{T_q}^{(n,p)}$. 
To set the scale, we use the relation 
$\sqrt{N_f/2}F_\pi/\sqrt{2}=246$ GeV. 
We again use the $F_\pi$ in the chiral limit.
To compare it with experiment, 
we use the cross section per nucleon ($\sigma_0$) 
instead of $\sigma_{SI}$.
The result is shown in Fig.~\ref{fig:sigma0}.
According to DM direct detection experiments\footnote{For a recent experiment, 
see e.g.\cite{Akerib:2013tjd}.}, 
our values for $\sigma_0$ are excluded,  
so that it may be difficult to explain 
the existence of DM as a techni-baryon.
However we note that there exist other contributions to the DM cross section,
e.g. gauge boson mediated interaction, and higher order terms, 
which might affect the DM cross section. 
Calculation of these contributions on the lattice is left for future investigations.

\vspace{-10mm}
\section{Summary}
The scalar mass and decay constant are very important parameters 
to probe a technicolor signature at the LHC. 
Based on the lattice theory,
we derived a relation between the scalar decay constant and 
the (flavor-singlet) scalar correlation functions. 
Our numerical result shows that 
the signal of the decay constant is as good as that of the mass. 
Although the accuracy of our result is not enough 
to precisely extrapolate towards the chiral limit, 
we obtained 
a rough estimate of the ratio $F_\sigma/F_\pi \sim 3$. 
This is useful for collider phenomenology.
Besides it, $F_\sigma$ can be also used for DM physics. 
We provided a DM effective theory based on the dilaton ChPT,
where a dilaton-DM coupling is related to $F_\sigma$ and the DM mass.
This coupling is constrained by DM direct detection experiments, 
e.g. LUX. 
We then discussed a possible scenario for techni-baryon DM. 
We note that all the results shown here are preliminary. 
Simulation at lighter 
fermion mass, and the direct computations on the lattice are needed 
to precisely estimate the above quantities.  

\vspace{5mm}
{\it Acknowledgments}
-- 
Numerical computations have been carried out on 
the high-performance computing systems at KMI ($\varphi$), 
at the Information Technology Center in Nagoya University (CX400), 
and at the Research Institute for Information Technology in Kyushu University (CX400 and HA8000). 
This work is supported by the JSPS Grant-in-Aid for Scientific Research (S) No.22224003, (C) No.23540300 (K.Y.), 
for Young Scientists (B) No.25800139 (H.O.) and No.25800138 (T.Y.), 
and also by the MEXT Grants-in-Aid for Scientific Research on Innovative Areas No.23105708 (T.Y.)
and No.25105011 (M.K.).
This work is supported by the JLDG constructed over the SINET of NII.
The work of H.O. is supported by the RIKEN Special Postdoctoral Researcher program.
E.R. acknowledges the support of the U.S. Department of Energy 
under Contract DE-AC52- 07NA27344 (LLNL).

\bibliographystyle{ws-procs975x65}
\bibliography{ws-pro-sample}

\end{document}